\documentstyle[prd,aps,eqsecnum,preprint,tighten,floats]{revtex}

\begin{document}
\draft

\input BoxedEPS
\SetTexturesEPSFSpecial
\HideDisplacementBoxes

\preprint{\vbox{\hfill OHSTPY--HEP--TH--95--034\\
          \vbox{\vskip0.5in} }}

\title{Light-Front QCD$_{1+1}$ Coupled to Chiral Adjoint Fermions}

\author{Stephen S. Pinsky and David G. Robertson}

\address{Department of Physics, The Ohio State University, Columbus,
OH 43210}

\date{\today}

\maketitle

\begin{abstract}
We consider $SU(N)$ gauge theory in 1+1 dimensions coupled to chiral
fermions in the adjoint representation of the gauge group.  With all
fields in the adjoint representation the gauge group is actually
$SU(N)/Z_N$, which possesses nontrivial topology.  In particular,
there are $N$ distinct topological sectors and the physical vacuum
state has a structure analogous to a $\theta$ vacuum.  We show how
this feature is realized in light-front quantization for the case
$N=2$, using discretization as an infrared regulator.  In the
discretized form of the theory the nontrivial vacuum structure is
associated with the zero momentum mode of the gauge field $A^+$.  We
find exact expressions for the degenerate vacuum states and the analog
of the $\theta$ vacuum.  The model also possess a condensate which we
calculate.  We discuss the difference between this chiral light-front
theory and the theories that have previously been considered in the
equal-time approach.

\end{abstract}

\pacs{ }

\section{Introduction}

The unique features of light-front quantization \cite{dir49} make it a
potentially powerful tool for the study of QCD.  Of primary importance
in this approach is the apparent simplicity of the vacuum state.
Indeed, naive kinematical arguments suggest that the physical vacuum
is trivial on the light front.  This cannot really be true, of course,
particularly in view of the important physics associated with the QCD
vacuum.  Thus it is crucial to understand the ways in which vacuum
structure can be manifested in light-front quantization.

There has recently been significant progress in this regard.  If one
uses discretization \cite{pab85} as an infrared regulator (i.e.,
imposes periodic or antiperiodic boundary conditions on some finite
interval in $x^-$), then any vacuum structure must necessarily be
connected with the $k^+=0$ Fourier modes of the fields.  Studies of
model field theories have shown that the zero modes can in fact
support certain kinds of vacuum structure; the long range phenomena of
spontaneous symmetry breaking \cite{ssb} as well as the topological
structure \cite{kap94,pin94} can in fact be reproduced with a careful
treatment of the zero mode(s) of the fields in a quantum field theory
defined in a finite spatial volume and quantized at equal light-front
time.

These phenomena are realized in quite different ways.  For example,
spontaneous breaking of $Z_2$ symmetry in $\phi^4_{1+1}$ occurs via a
{\em constrained} zero mode of the scalar field \cite{czm}.  There the
zero mode satisfies a nonlinear constraint equation that relates it to
the dynamical modes in the problem.  At the critical coupling a
bifurcation of the solution occurs.  These solutions in turn lead to
new operators in the Hamiltonian which break the $Z_2$ symmetry at and
beyond the critical coupling.  Quite separately, a {\em dynamical}
zero mode was shown in Ref. \cite{kap94} to arise in pure $SU(2)$
Yang-Mills theory in 1+1 dimensions.  A complete fixing of the gauge
leaves the theory with one degree of freedom, the zero mode of the
vector potential $A^+$. The theory has a discrete spectrum of
zero-$P^+$ states corresponding to modes of the flux loop around the
finite space.  Only one state has a zero eigenvalue of the energy
$P^-$, and is the true ground state of the theory. The nonzero
eigenvalues are proportional to the length of the spatial box,
consistent with the flux loop picture.  This is a direct result of the
topology of the space. As the theory considered there was a purely
topological field theory, the exact solution was identical to that in
the conventional equal-time approach on the analogous spatial topology
\cite{het93}.

In the present work we shall consider QCD$_{1+1}$ coupled to adjoint
fermionic matter. The theory has previously been studied in the limit
of large $N_c$ \cite{koz95}, for $N_c=2$ at finite temperature
\cite{smilga}, and for 2 and 3 colors in equal-time quantization in
the small-volume limit \cite{les94}.  This model also arises in the
context of supersymmetry \cite{kut94}.  It is interesting for our
purposes because it possesses a vacuum structure analogous to a
$\theta$ vacuum.  As first shown in Ref. \cite{witten}, for $SU(N)$
gauge fields the vacuum has a $Z_N$ topological structure.
Furthermore, for $N=2$ there is a nonvanishing bilinear condensate
\cite{smilga}.  Our goal is to see to what extent this structure can
be demonstrated on the light front.

Similar theories coupled to adjoint scalars have also been studied
recently \cite{dek93}.  Here the scalar field can be thought of as the
$k_\perp=0$ remnant of the transverse gluon component in QCD$_{2+1}$.
The study of these theories is part of a long-term program to attack
QCD$_{3+1}$ through the zero mode sectors starting with studies of
lower dimensional theories which are themselves zero mode sectors of
higher dimensional theories \cite{kap93,kap94b}. A complete gauge
fixing has recently been given for QED$_{3+1}$ which further supports
this program \cite{kar94}.  In all of these cases, the central problem
was to disentangle the dependent from the independent fields in the
context of a particular gauge fixing.

For massless fermions in 1+1 dimensions, the usual light-front
approach corresponds to a theory with only a single chirality of
fermion; only the right-handed particles are present
\cite{mccartor88}.  Thus the light-front model we consider is
not the same as the the model discussed in Ref. \cite{les94}, which
contains both left- and right-handed dynamical fermions.  By not
including the left-handed fields we are essentially considering a
chiral version of the theory discussed in Ref. \cite{les94}.  This
implies that any condensate that we will find will be fundamentally
different in structure.  In order to compare with with the theory
discussed in Ref. \cite{les94} we would need to include the additional
structure of the left-handed fields.  We hope to report on this in the
near future.

The topological classification of the vacua is unaffected by the
chiral nature of the theory, however, so we still expect to find $N$
degenerate vacua for $SU(N)$ gauge fields.  For the case $N=2$
considered here we shall indeed find two vacuum states.  As suggested
above, the physics of these states is closely related to the only zero
mode present in the theory, that of $A^+$.  The properties of this
mode, in turn, are tied up with issues of gauge fixing, Gribov
horizons, etc.

The remainder of the paper is organized as follows.  In the next
section we define the theory, including the gauge fixing, and outline
our calculational scheme.  Section 3 is devoted to the definition of
the currents and charge operators, which are important for defining a
suitable physical subspace.  Next we study the vacuum sector of the
theory and find the two ground states.  In Section 4 we discuss the
condensate and obtain an exact expression for it.  Section 5 contains
some discussion and directions for future work.

\section{Definition of the Theory and Gauge Fixing}

We consider an $SU(N)$ gauge field coupled to adjoint
fermions in one space and one time dimension.  Since all fields
transform according to the adjoint representation, gauge
transformations that differ by an element of the center of the group
actually represent the same transformation and so should be
identified.  Thus the gauge group of the theory is $SU(N)/Z_N$, which
has nontrivial topology: $\Pi_1[SU(N)/Z_N]=Z_N$, so that one expects
$N$ topological sectors.  This situation differs from the case when
the matter fields are in the fundamental representation, where the
gauge group is $SU(N)$ and the first homotopy group is trivial.

The Lagrangian for the theory is
\begin{equation}
{\cal L} = - {1 \over 2} Tr (F^{\mu \nu} F_{\mu \nu}) + {1 \over 2}
Tr(\bar\psi\gamma ^{\mu} \buildrel \leftrightarrow \over
D_{\mu}\psi)\; ,
\end{equation}
where $D_{\mu} = \partial_{\mu} + ig [ A_\mu,\ ] $ and $F_{\mu \nu} =
\partial_{\mu}A_{\nu} - \partial_{\nu} A_{\mu} + ig [A_{\mu}, A_{\nu}
]$.  We employ light-front quantization, defining $x^{\pm} = (x^0 \pm
x^1)/ \sqrt 2$ and taking $x^{+}$ to be the evolution parameter.  A
convenient representation of the gamma matrices is $\gamma ^0 =
\sigma^2$ and $\gamma^1 = i\sigma^1$, where $\sigma^a$ are the Pauli
matrices.  With this choice, the Fermi field may be taken to be
hermitian.  It is natural in light-front quantization to break the
Fermi field into two components
\begin{equation}
\Psi _\pm\equiv{1\over2}\gamma^0\gamma^\pm\Psi\; ;
\end{equation}
in two dimensions (only) these are the same as chiral projections, so
that
\begin{equation}
{1\over2}\gamma^0\gamma^+ \Psi = {\Psi _R \choose 0} ,\quad
{1\over2}\gamma^0\gamma^- \Psi = {0 \choose \Psi _L} .
\label{project}
\end{equation}
When the fermions are massive, $\Psi_+$ appears to be the only
independent degree of freedom.  For the massless theory considered
here, both $\Psi_+$ and $\Psi_-$ must be considered to be independent
fields \cite{mccartor88}.

We shall focus on the case $N=2$, in which case the matrix
representation of the fields makes use of the $SU(2)$ generators
$\tau^a=\sigma^a/2$.  It is convenient to introduce a color helicity
basis, defined by $\tau^\pm\equiv\tau^1\pm i\tau^2$ with $\tau^3$
unchanged.  These satisfy $[\tau^+, \tau^- ] = \tau^3$ and $[\tau^3 ,
\tau^{\pm}] = \pm\tau^{\pm}$.  In terms of this basis the
matrix-valued fields are given by, for example,
\begin{equation}
A^{\mu} = A _3 ^{\mu} \tau ^3 + A ^{\mu} _+ \tau^+ + A_-^{\mu}\tau^-\; ,
\end{equation}
where $A^\mu_\pm\equiv A^\mu_1\pm iA^\mu_2$.  (Note that
$(A^\mu_+)^\dagger=A^\mu_-$.)  The Fermi field will be similarly
written as
\begin{equation}
\Psi _{R/L} = \psi _{R/L} \tau ^3 + \phi _{R/L} \tau ^+ +
\phi^\dagger _{R/L} \tau ^-\; ,
\label{helicity}
\end{equation}
with $\phi_{R/L}\equiv \Psi^1_{R/L}+i\Psi^2_{R/L}$.  Under a gauge
transformation the gauge field transforms in the usual way and the
Fermi field transforms according to
\begin{equation}
\Psi _{R/L} \Rightarrow U \Psi _{R/L} U^{-1}\; ,
\end{equation}
where $U$ is a spacetime-dependent element of $SU(2)$.

We shall regulate the theory by putting it in a light-front spatial
box, $-L < x^{-} < L$, and imposing periodic boundary conditions for
the gluon fields $A^{\mu}$ and anti-periodic boundary conditions for
the Fermi field.  In this approach, the subtle aspects of formulating
the model have to do with the zero-momentum modes of the fields.  It
is here, also, that any nontrivial vacuum structure must reside.

In the present model the subtlety is in fixing the gauge.  It is most
convenient in light-front field theory to choose the light-cone gauge
$A^+=0$.  Here, however, since the gauge transformation must be
periodic up to an element of the center of the gauge group (here
$Z_2$), we cannot gauge the zero mode of $A^+$ to zero \cite{frn81}.
Thus we choose $\partial_-A^+=0$.  We can make a further global (i.e.,
$x^-$-independent) rotation so that the zero mode of $A^+$ has only a
color 3 component,
\begin{equation}
A^+ = v (x^+) \tau ^3 \equiv V(x^+)\; ,
\end{equation}
and simultaneously rotate $A^-$ so that it has no color 3 zero mode
\cite{kap94}.

At this stage the only remaining gauge freedom involves certain
``large'' gauge transformations, which we shall denote $T_n$.  This
freedom is best studied in terms of the dimensionless variable $z = g
v L/ \pi$, which $T_n$ shifts by an integer:
\begin{equation}
T_nz T^{-1}_n = z + n\; .
\label{whattdoes}
\end{equation}
In addition, $T_n$ generates a space-dependent phase rotation on the
matter field $\phi _{R/L}$
\begin{equation}
T_n \phi _{R/L} T^{-1}_n = e ^{-in\pi x ^-/ L} \phi _{R/L}\; ,
\label{tsymm}
\end{equation}
which however preserves the anti-periodic boundary condition on
$\phi_{R/L}$.  This gauge freedom is an example of the Gribov
ambiguity \cite{gri78}.  We can use it to bring $z$ to a finite
domain, for example $0 < z < 1$ or $-1 < z < 0$.  Once this is done
all gauge freedom has been exhausted and the gauge fixing is
completed.  Only physical degrees of freedom remain.

After gauge fixing $T_n$ is no longer a symmetry of the theory, but
there is a symmetry of the gauge-fixed theory that is conveniently
studied by combining $T_1$ with the so-called Weyl symmetry, denoted
by $R$.  Under $R$,
\begin{equation}
RzR^{-1} = -z \quad {\rm and} \quad R \phi_{R/L} R^{-1} =
\phi^\dagger_{R/L}\; .
\label{rsymm}
\end{equation}
This is also not a symmetry of the gauge-fixed theory, as it takes $z$
out of the fundamental domain.  The symmetry $T_1R$, however, which is
closely related to charge conjugation, plays an important role in the
gauge-fixed theory as will be discussed in detail below.

The Hamiltonian $P^-$ takes a very standard form
\begin{equation}
P^- = g\int _{-L} ^L dx^- Tr\left(AJ^+\right)
+ 2 L \partial _+ v \partial _+v \; ,
\label{basicham}
\end{equation}
where $A \equiv A^-$ and $J^+ = {1\over\sqrt 2}[\Psi _{R},\Psi_R]$.
The field $A$ is nondynamical and is obtained by solving Gauss law,
\begin{equation}
-D_- ^2 A = g J^+\; .
\end{equation}
Resolving this into its color components we have
\begin{eqnarray}
-\partial {^2 _-} A_3 &=& gJ_3^+ \\
-(\partial _- + igv)^2 A_+ &=& g J_+^+ \\
-(\partial _- - igv)^2 A_- &=& g J_-^+\; .
\end{eqnarray}
The first of these can be solved for the normal mode part of $A_3$
(recall that the zero mode has been gauged away).  Because of the
boundary conditions and the restriction of $v$ to a finite domain, the
covariant derivatives appearing in the second and third equations have
no zero eigenvalues.  Thus they can be inverted to solve for $A_+$ and
$A_-$.  The only part of Gauss' law that remains to be implemented is
the zero mode of the first equation, which reduces to the vanishing of
the zero mode of $J^+_3$.  This condition must be imposed on the
states and defines the physical subspace of the theory:
\begin{equation}
Q_3 \vert {\rm phys} \rangle = 0\; ,
\end{equation}
where
\begin{equation}
Q_3 = \int _{-L} ^L dx ^- J_3^+ \; .
\end{equation}

After implementing the solution of Gauss' law we have
\begin{equation}
P^- = {g^2 \over 2(2 \pi )^2 }\pi^2_z - g^2 \int
^L _{-L} dx^- Tr \left( J^+ {1\over D_-^2 } J^+ \right)\; ,
\label{hamilton}
\end{equation}
where $\pi _z$ is the momentum conjugate to the quantum mechanical
degree of freedom $z = gvL/\pi$, defined so that $[z,\pi_z] = i$.  In
this form it is clear that the dynamical variables are $\Psi_R$ and
$z$.  We shall use a Fock space representation for the Fermi degrees
of freedom and a Schr\"odinger representation for the $z$ degree of
freedom. Thus states will be written as tensor products of the general
form $\psi(z)\otimes |{\rm Fock}\rangle $, and $\pi_z$ will be
represented as a derivative operator: $\pi_z=-i\partial_z$.  The
Fourier expansion of $\Psi_R$ has the usual form
\begin{eqnarray}
\psi _R &=& {1 \over 2 ^{1/4} \sqrt {2L}} {\sum_n}
\left(a_n e^{-ik {^+ _n} x^-} + a{_n ^\dagger}
e^{ik{_n ^+}x^-} \right) \\
\phi _R &=& {1 \over 2 ^{1/4} \sqrt {2L}} {\sum_n}
\left(b_n e^{-ik {^+ _n} x^-} + d{_n ^\dagger}
e^{ik{_n ^+}x^-} \right)\; ,
\end{eqnarray}
where the sums are over the positive half odd integers and $k{^+ _n} =
n \pi /L$.  The Fock operators obey the standard commutation relations
\begin{equation}
\{a{^\dagger _n}, a_m \} = \{ b{_n ^\dagger}, b_m \} = \{ d{_n
^\dagger}, d_m \} = \delta _{n, m}
\end{equation}
These result in the Heisenberg equation correctly reproducing the
equation of motion for $\Psi_R$,
\begin{equation}
D_+ \Psi _R = \partial _+ \Psi _R + ig [A, \Psi _R ] = 0\; .
\end{equation}
In addition, of course, $[z,\Psi_R]=[\pi_z,\Psi_R]=0$.

\section{Current Operators and the Physical Subspace}

Next let us discuss the definition of the current $J^+$ and the
associated charge operator in more detail.  The relation
\begin{equation}
J^+ = {1 \over \sqrt 2} [\Psi_R , \Psi _R]
\end{equation}
is ill-defined as it stands.  We shall regulate it using a gauge
invariant point splitting:
\begin{equation}
J_3 ^+ = \lim _{\epsilon \to 0}
\left[ e^{-ig \int {_{x^- - \epsilon } ^{x^-}} v\tau^3 dx^-}
\Psi_R (x ^- -\epsilon )
e^{ig \int {^{x^-} _{x^- -\epsilon ^-}} v\tau^3dx^-} ,\Psi_R (x^-)
\right]\; .
\label{ham}
\end{equation}
We find that the current $J^+$ acquires a gauge correction
\begin{equation}
J^+ = {J'}^+ + {g \over 2\pi}v (x^+) \tau ^3\; ,
\label{anomaly}
\end{equation}
where ${J'}^+$ is the naive normal-ordered current.  This result is
potentially upsetting, as the charge calculated from this current
would seem to have an unwanted time dependence.  Note, however, that
any $x^-$-independent piece of the current $J^+_3$ couples to the zero
mode of $A_3$, which has been gauged to zero [see
Eqn. (\ref{basicham})].  Therefore what enters the dynamics is not the
full current but the current with the ``anomaly'' (and any other color
3 zero mode) removed.  It follows that ${J'}^+$ is what will appear in
the equations of motion. In particular, Gauss' Law takes the form
\begin{equation}
-D_-^2 A = g {J'}^+\; .
\end{equation}

Since the zero mode of $J^+_3$ does not appear in the dynamics of the
theory, one can ask how it is to be defined.  As we have seen, it is
necessary to define the current in such a way that its zero mode has
no gauge correction.  The presence of such a term would be quite
unpleasant as the charge, which is supposed to project out physical
states, would be time-dependent.  Another property that the charge
should possess it $T_1R$ symmetry.  In order to discuss this it is
helpful to consider the transformation properties of the fields and
the naive charge
\begin{equation}
Q'_3 = {\sum_n} (d_n ^\dagger d_n - b _n ^\dagger b_n )
\end{equation}
under $T_1$ and $R$.  From Eqn. (\ref{tsymm}) we see that $T_1$ gives
rise to a spectral flow,
\begin{eqnarray}
T_1 b_n T^{-1}_1 & = & b_{n-1}\; , \quad n > 1/2 \\
T_1 d_n T^{-1}_1 & = & d _{n+1} \\
T_1 b _{1/2} T^{-1}_1 & = & d {_{1/2} ^\dagger}\; .
\end{eqnarray}
This leads to
\begin{equation}
T_1 Q'_3 T^{-1}_1 = Q' _3 -1\; .
\end{equation}
In addition, $T_1$ shifts $z$ by unity [Eqn. (\ref{whattdoes})].
Under $R$ symmetry, meanwhile, we find from Eqn. (\ref{rsymm}) that
\begin{equation}
R b_n R^{-1} = d_n\; ,
\end{equation}
which gives
\begin{equation}
R Q'_3 R^{-1} = -Q' _3\; .
\end{equation}
Its action on $z$ is to take $z\Rightarrow -z$ [Eqn. (\ref{rsymm})].
Putting these together we find
\begin{equation}
T_1R Q'_3 R^{-1}T_1^{-1} = 1-Q'_3
\end{equation}
and
\begin{equation}
T_1R z R^{-1}T_1^{-1} = -z-1\; .
\end{equation}
This represents a symmetry of the theory since it maps the fundamental
domain $-1<z<0$ onto itself.  In fact, $T_1R$ represents a reflection
of the fundamental domain about its midpoint $z=-1/2$, coupled with a
spectral flow of the fermionic degrees of freedom.  It is
straightforward to check that the Hamiltonian Eqn. (\ref{hamilton})
commutes with $T_1R$.

Now the charge operator we use to select the physical subspace must
also be invariant under $T_1R$, so that the physical subspace is
mapped into itself under the transformation.  Clearly, $Q'_3$ is not
invariant and so cannot be used for this purpose.  Note, however, that
{\em two} applications of the transformation $T_1R$ leave $Q'_3$, as
well as the fundamental domain, invariant.  Thus if we define the
physical subspace to consist of all states annihilated either by
$Q'_3$ or by $1-Q'_3$, then it will be invariant under the $T_1R$
transformation and this will represent a true symmetry of the theory.
As this has all the properties we require, we shall adopt it as the
definition of the physical subspace.  Note that it is stable under
time evolution, since $[Q'_3,P^-]=0$.

\section{ Vacuum States of The Theory }

The Fock state containing no particles will be called $\vert
V_0\rangle$.  If it is one of a set of states that are related to one
another by $T_1$ transformations, and which will be denoted
$|V_M\rangle$, with $M$ any integer.  These are defined by
\begin{equation}
|V_M\rangle\equiv (T_1)^M|V_0\rangle\; ,
\end{equation}
where $(T_1)^{-1}=T_{-1}$.  It is straightforward to determine the
particle content of the $|V_M\rangle$.  Consider, for example, the
$T_1$ transform of
\begin{equation}
b {^\dagger _{1/2}} b_{1/2} \vert V_0 \rangle = 0\; ,
\end{equation}
which is
\begin{equation}
T_1 b{^\dagger _ {1/2}} T^{-1}_1 T_1 b_{1/2} T^{-1}_1 T_1 \vert V_0
\rangle = 0\; .
\end{equation}
Using Eqn. (\ref{tsymm}) we have
\begin{equation}
d_{1/2} d{^\dagger _{1/2}} \vert V_1 \rangle = 0
\end{equation}
which implies
\begin{equation}
d{^\dagger _{1/2}} d_{1/2} \vert V_1 \rangle = \vert V_1 \rangle
\end{equation}
and therefore $\vert V_1 \rangle$ will have one $d_{1/2}$ background
particle. One can show that $\vert V_1 \rangle$ has no other content;
$\vert V_1 \rangle \equiv \vert 0; 0; 1/2 \rangle \equiv d{^\dagger
_{1/2}} \vert 0; 0; 0 \rangle$, using the Fock space notation $\vert
\{ n_a \}; \{ n _b \}; \{ n_d \} \rangle $.  Under the $R$
transformation $d \rightarrow b$, so that
\begin{equation}
R \vert V_1 \rangle = \vert V_{-1} \rangle \equiv \vert 0 ; 1/2 ; 0
\rangle \; .
\end{equation}
Similar relations hold for the state $\vert V_M\rangle $ where
$-\infty < M < \infty$ and $M < 0$ correspond to states with
background $b$ particles.

These state are related by gauge transformations and are therefore
``physically equivalent," but for different values of $z$, since $T_1$
shifts $z$ by unity.  In a given domain, for example $-1 < z < 0$,
these state are to be considered as inequivalent.  Note that only
$|V_0\rangle$ and $|V_1\rangle$ are in the physical subspace as we
have defined it; the first is annihilated by $Q'_3$ while the second
is annihilated by $1-Q'_3$.

As discussed previously, we shall use a Schr\"odinger representation
for the gauge degree of freedom described by $z$ and $\pi_z$.  In this
mixed representation, states are written in the form
\begin{equation}
\psi (z) \vert \{ n_a \}; \{ n _b \}; \{ n_d \} \rangle\; .
\end{equation}
The object is now to find the lowest-lying eigenstates of the
Hamiltonian $P^-$ which are linear combinations of states of this
form.

The Hamiltonian is given by
\begin{equation}
P^- = -{g^2 \over 2(2 \pi )^2} {d^2 \over dz^2} -g^2
\int _{-L} ^L dx^- Tr \left( J ^+ {1 \over D_-^2} J^+\right)\; .
\label{finalham}
\end{equation}
It is convenient to separate $P^-$ into a ``free'' part and an
interaction,
\begin{equation}
P^- = P{_0 ^-} + P_I^-\; .
\end{equation}
$P_0^-$ includes all $z$-dependent $c$-numbers and one-body Fock
operators that arise from normal ordering Eqn. (\ref{finalham}), and
has the form
\begin{equation}
P{_0 ^-} = C(z) + V (z)\; ,
\end{equation}
where $C(z)$ is a $c$-number function of $z$ and
\begin{equation}
V(z) = {\sum_n} \left( A_n (z) a{_n ^\dagger} a_n + B_n (z) b{_n
^\dagger} b_n + D_n (z) d{_n ^\dagger} d_n \right)\; .
\end{equation}
The explicit forms of $C(z)$, $A_n (z)$, $B_n (z)$, and $D_n (z)$ are
given in the Appendix.  $P_I^-$ is a normal-ordered two body
interaction.  We do not display it here as it is unnecessary for our
present purposes.  Note that $P_0^-$ itself is invariant under $T_1$
and $R$:
\begin{equation}
T_1P{_0 ^-} T^{-1}_1 = P{_0 ^-} (z) \quad , \quad R P{_0 ^-} R ^{-1} =
P{_0 ^-}\; .
\end{equation}
This is not true of $C(z)$ and $V(z)$ individually.

\begin{figure}
\centering\BoxedEPSF{pot.epsf}
\caption{\label{ff1} The potential $C(z)$ as a function of $z=
gvL/2\pi$, and the associated wavefunction}
\end{figure}

Consider a possible vacuum state $\zeta (z) \vert V_0 \rangle$, where
we choose the fundamental domain $-1 < z < 0$.  We consider a matrix
element of $P^-$ acting between this state and an arbitrary Fock
state.  The only non-vanishing matrix element is
\begin{equation}
\langle V_0 \vert P^- \zeta (z) \vert V_0 \rangle = \epsilon _0 \zeta(z)
\end{equation}
which leads to Schr\"odinger equation for $\zeta (z)$:
\begin{equation}
\Biggl [ -{g^2 \over (2\pi )^2} {d^2 \over dz^2} + {g^2 \over 2} C(z)
\Biggl ] \zeta (z) = \epsilon _0 \zeta (z)\; .
\end{equation}
The ``potential'' $C(z)$ is shown in Figure 1 and has a minimum at $z
= 0$.  It is straightforward to solve this quantum mechanics problem
with the boundary conditions $\zeta(0) =0$ and $\zeta (-1) = 0$.
These boundary conditions are the result of a number of studies
\cite{kap94,het93} of the behavior of states at the boundaries of
Gribov regions (in our case, the integer values of $z$). The shape of
the wave function is shown in Figure 1.  To discuss the symmetries of
this theory we will find it convenient to define $\zeta (z)$ outside
of the fundamental domain.  We shall define it to be symmetric about
$z = 0$ since $C(z)$ is symmetric about $z=0$.

Now let us consider the state $\bar \zeta (z)\vert V_1 \rangle$.
Projecting the matrix element of $P^-\bar\zeta (z) \vert V_1 \rangle$
with $\vert V_1 \rangle $ we find
\begin{equation}
\Biggl [ -{g^2 \over (2 \pi)^2} {d ^2 \over dz^2} + {g ^2 \over 2}
\bigl [ C(z) + D_{1/2} (z) \bigr ] \biggr ] \bar \zeta (z) = \epsilon
_1 \bar \zeta (z)\; .
\end{equation}
{}From the explicit forms of $C(z)$ and $D _{1 / 2} (z)$ given in the
Appendix it can be shown that $C(z) + D_{1/2} (z) = C(z+1)$.
This is of course just the realization of the $T_1$ invariance of $P
^- _0$.  Setting $\bar \zeta (z) \equiv \zeta (z+1)$ we find the
$\epsilon _1 \equiv\epsilon _0$ and the Schr\"odinger equation is
\begin{equation}
\Biggl [ {-g^2 \over (2 \pi )^2} {d^2 \over dz^2} + {g^2 \over 2}
C(z+1)\Biggr ]\zeta (z+1) =\epsilon _0 \zeta (z+1)\; .
\end{equation}
The functions $C(z+1)$, $\zeta (z+1)$, $C(z)$ and $\zeta (z)$ are
shown in Figure 2.  From this figure it is clear that $\zeta (z) \vert
V_0 \rangle $ and $\zeta (z + 1) \vert V_1 \rangle $ are two
degenerate vacuum states in the domain $ -1 < z < 0 $.
\begin{figure}
\centering\BoxedEPSF{pot2.epsf}
\caption{\label{ff2}
The potentials $C(z)$ and $C(z+1)$ in the fundamental modular domain
and the associated wave
functions}
\end{figure}
We have found what we believe are the two expected degenerate vacua
for this theory in the domain $-1 < z < 0$. From Figure 2 we see that
these state are reminiscent of the equal-time result
\cite{les94}. There are some significant differences with the
equal-time results however.  On the light-front we have an exact
expression for the vacuum states, while in the equal-time approach
this could only be determined approximately.

The action of the operator $T_1R$ on one of the states $|V_M\rangle$
has strictly speaking only been defined up to a phase,
\begin{equation}
T_1R|V_M\rangle=e^{i\theta_M}|V_{1-M}\rangle\; .
\end{equation}
The phase $\theta_M$ is arbitrary, except that it must satisfy
\begin{equation}
e^{i\theta_M}e^{i\theta_{1-M}}=1\; ,
\end{equation}
which follows from $(T_1R)^2=1$.  We can now construct a vacuum state
that is phase-invariant under the symmetry $T_1R$ by superposing our
two ``$n$-vacua'':
\begin{equation}
\vert \theta \rangle ={1 \over \sqrt{2}}[ \zeta (z) \vert V_0 \rangle
+ e ^{i\theta_0} \zeta (z + 1) \vert V_1 \rangle]\; .
\label{vacuum}
\end{equation}
where $\zeta(z)$ is normalized to one.  It is typically necessary to
build the theory on such a vacuum state in order to satisfy the
requirements of cluster decomposition.

Finally we would like to briefly discuss $P^+$.  We use an explicitly
gauge-invariant form of $P^+$
\begin{equation}
P^+ = - i \sqrt2 \int _{-L}^L dx^- Tr\left(\Psi _R D_- \Psi
_R\right)\; .
\end{equation}
This is a singular operator and requires regularization and
renormalization.

We have done this in two ways, using the gauge-invariant
point-splitting discussed earlier and also a $\zeta$-function
regularization.  Both procedures give the same result:
\begin{equation}
P^+ = {\pi \over L } {\sum_n} n (a{_n ^\dagger} a_n + b{_n^\dagger}
b_n + d{_n ^\dagger}d_n ) + {\pi \over L} z Q{_3 ^{\prime}} + {\pi
\over 2L} z^2\; .
\end{equation}
One can explicitly show that this expression is $T_1$ and $R$
invariant (of course, the exact form of the non-standard terms is
essential for this result).  The Poincar\'e algebra here is
essentially $[P^-, P^+ ] = 0$, Explicit calculation gives $[P^-, P^+]
= {i \pi \over L} \Bigl (2 \pi _z Q' _3 - (z \pi _z + \pi _z z ) \Bigr
)$.  However the matrix element of the commutator with all physical
states vanish. Thus the Poincar\'e algebra is valid in physical
subspace.  This result rests on the fact that we only use the ground
state wave function $\zeta(z)$ to construct physical states.  There
are higher-energy solutions to the quantum mechanics problem in $z$;
however, the energy differences are proportional to $L$ since these
energy levels are associate with quantized flux loops that circulate
around the closed $x^-$ space.  The spectrum of states associated with
these very high-energy states decouple in the continuum limit and can
be ignored.

\section{ The Condensate }

It is generally accepted that QCD in 1+1 dimensions coupled to adjoint
fermions develops a condensate. So far this condensate has only been
calculated in various approximations. It has been calculated in the
large-$N_c$ limit in Ref. \cite{koz95} and in the small-volume limit
for $SU(2)$ in Ref. \cite{les94}. The theory we are considering here
is a chiral theory with only right-handed fermions. It is natural to
consider such a theory in a light-front quantized theory because the
light-front projections Eqn. (\ref{project}) naturally separate the
left- and right-handed parts of the theory.  Since the theory
considered in Refs. \cite{koz95,les94} has both dynamical left- and
right-handed fields we do not expect to obtain the same result as
those calculations.

The two vacuum states $\zeta(z) \vert V_0 \rangle$ and $\zeta(z + 1)
\vert V_1 \rangle$ of our chiral theory are both exact ground states.
Since we only have right-handed dynamical fermions we only have a
spectral flow associated with the right-handed operators, and thus the
two physical spaces in the fundamental domain therefore differ by a
single fermion. They effectively block diagonalize $P^-$ into two
noncommunicating sectors. One sector is built on a vacuum with no
background particles and the other built on a state with one
background particle.  Therefore the matrix element of any color
singlet operator between these two sectors are expected to vanish.
While this theory will not generate a fundamental color-singlet
condensate, it does develop a vacuum expectation value for the fermion
field.

It is straightforward to calculate the vacuum expectation value from
Eqn. (\ref{vacuum}):
\begin{equation}
\int _{-L}^L dx^- \langle \theta \vert \; \phi_R(x^-) \; \vert \theta
\rangle = e^{-i\theta} \; {\sqrt{L} \over \pi} \int_{-1}^0 dz
\zeta(z+1) \zeta(z)\; .
\label{condensate}
\end{equation}
Since Eqn. (\ref{vacuum}) is an exact expression for the vacuum,
Eqn. (\ref{condensate}) is an exact expression for the vacuum
expectation value.

The equal-time calculation by contrast has both dynamical right-handed
and left-handed particles and a spectral flow associated with both.
Therefore its two physical spaces for $SU(2)$ would differ by two
fermions, which in turn gives rise to a color-singlet condensate.

\section{ Conclusions }

We have shown that in QCD coupled to chiral adjoint fermions in two
dimensions the light-front vacuum is two-fold degenerate as one would
expect on general grounds.  The source of this degeneracy is quite
simple.  Because of the existence of Gribov copies, the one gauge
degree of freedom, the zero mode of $A^+$, must be restricted to a
finite domain. The domains of this variable, which after normalization
we call $z$, are bounded by the integers.  Furthermore there is a
symmetry of the theory under reflections about the midpoint of the
fundamental domain.  Thus the potential of the vacuum state in the
variable $z$ can either have a minimum at $z=1/2$ or have multiple
minima.  It has recently been seen that for adjoint scalars the
minimum is at $z=1/2$.  In the problem with adjoint fermions described
here there are two minima at the ends of the domain.

In the light-front formalism we obtain an exact expression for the
vacuum states and we can solve for their fermionic content exactly.
We find that these states are very different.  They differ because the
$T_1$ transformation gives rise to a spectral flow for the
right-handed fermion; thus the two vacuum states differ in the
background fermion number and color that each carries.  We form the
analog of a $\theta$ vacuum from these two-fold degenerate vacuum
states which respects all of the symmetries of the theory.  We find
that field $\phi_R$ has a vacuum expectation value with respect to
this $\theta$ vacuum and we find a exact expression for this vacuum
expectation value. This chiral theory differs from the theories that
have been studied in the equal time formulation \cite{koz95,les94},
because the equal time theory has both dynamical left- and
right-handed fields.  We expect that if we were to couple together two
light-front theories, one with dynamical right-handed particles and
the other with dynamical left-handed particles, then the resulting
degenerate vacuum and condensate would be similar to those discussed
in the equal-time theory.  We shall discuss the details of such a
theory elsewhere.

\acknowledgments
\noindent
It is a pleasure to thank K. Harada for many helpful discussions.
S.S.P. would like to thank the Max-Planck Institut f\"ur Kernphysik,
Heidelberg, for their hospitality during portions of this work.  This
work was supported in part by a grant from the US Department of
Energy.  In addition, D.G.R. was supported during the early stages of
this work by the National Science Foundaton under Grants Nos.
PHY-9203145, PHY-9258270, and PHY-9207889.  Travel support was
provided in part by a NATO Collaborative Grant.

\appendix

\section{The Gauge Potential}

We list here the explicit forms of the functions $C$, $A_n$, $B_n$,
and $D_n$ discussed in Sect. 4.  The $c$-number function $C(z)$ must
be retained here since it is an operator in $z$ space. The divergence
is easily seen to be a true constant and therefore can be subtracted.
\begin{equation}
C(z) = {1 \over 4}{ \sum _{n,m}} \Biggl [ {1 \over (n + m + z )^2} +
{1 \over (m + n - z )^2} \Biggr ] - {1 \over 2} \sum _{p = 1}
^{\infty} {1 \over P}
\end{equation}
\begin{equation}
A_n = {{1 \over 4} \sum _m} \Biggl [ {1 \over (n - m - z)^2} +
     {1 \over (n - m + z )^2} - {1 \over (n - m + z )^2} -
     {1 \over (n + m - z )^2 } \Biggr ]
\end{equation}
\begin{equation}
B_n = {1 \over 4} {\sum _m} \Biggl [ {1 \over (m - n + z
)^2 } - {1 \over (n + m - z )^2 } \Biggr ]
\end{equation}
\begin{equation}
D_n = \sum _m \Biggl [{1 \over (m - n - z)^2 } - { 1 \over (n +
m + z )^2} \Biggr ]
\end{equation}
%


\end{document}